\documentclass[a4paper,11pt]{article}
\usepackage{pos}

\title{Electromagnetic form factors of baryons \\ in nuclear medium}
\ShortTitle{Electromagnetic form factors of baryons in nuclear medium}

\author*{G.~Ramalho}
\author{K.~Tsushima}
\author{J.~P.~B.~C.~de Melo}

\affiliation{Laborat\'orio de 
F\'{i}sica Te\'orica e Computacional -- LFTC, \\
Universidade Cruzeiro do Sul and Universidade Cidade de  S\~ao Paulo,  \\
01506-000,   S\~ao Paulo, SP, Brazil}

\emailAdd{gilberto.ramalho2013@gmail.com}

\abstract{The electromagnetic structure of the baryons is modified 
in the nuclear medium. 
The modifications can be inferred from the comparison 
between the electromagnetic form factors in medium with 
the respective form factor in vacuum.
Of particular interest is the ratio between 
the electric and magnetic form factors in medium ($G_E^*/G_M^*$) 
and vacuum ($G_E/G_M$) of the octet baryon.
The deviation of the double ratios ($G_E^*/G_M^*)/(G_E/G_M)$ from
unity measures the impact of the medium modification 
of the electromagnetic structure in a nuclear medium.
Measurements of the double ratios $(G_E^*/G_M^*)/(G_E/G_M)$ 
for different nuclear densities may become available 
in a near future using the polarization-transfer method 
developed at Jefferson Lab.
We present estimates of the double ratios of octet baryons
based on a covariant constituent quark model,
which takes into account pion cloud excitations of the baryon cores,
for different nuclear densities.
Our results manifest different features, namely, 
enhancement or quenching depending on the baryon flavor content.}

\FullConference{%
  *** Particles and Nuclei International Conference - PANIC2021 ***\\
  *** 5 - 10 September, 2021 ***\\
  *** Online ***
}

\begin{document}
\maketitle

\section{Introduction}

Since the discovery of the EMC effect~\cite{Saito07}
it has been known that the hadrons change their  
properties in a nuclear medium~\cite{MediumDR,Medium1}.
The natural interpretation is that a strong mean field
modifies the properties of the degrees of freedom of QCD, 
the quarks and gluons. 
In these conditions, we can expect that 
the electromagnetic form factors of the baryons
are modified compared to the vacuum and 
that the impact of the changes increase with 
the nuclear density $\rho$.

The modification due to the nuclear medium 
can be quantified comparing the measured form factors in medium 
with the measured form factors in vacuum.
The open question is: can we measure the electric and magnetic 
form factors of baryons in medium, 
when measurements in vacuum are at the moment restricted to 
the nucleon (proton and neutron).
The answer is, the individual form factors cannot be measured directly, 
but, with the polarization-transfer method, it is 
possible to measure the ratio $G_E/G_M$ in vacuum 
and in medium~\cite{Ndata,Dieterich01}.
To represent the electric and magnetic form factors 
in medium we use $G_E^\ast$ and $G_M^\ast$, respectively.

The polarization-transfer method has been applied 
in the Jefferson Lab to measure
the ratio $G_E/G_M$ of the proton and neutron in vacuum~\cite{Ndata}.
In the experiments the nucleon is scattered by 
polarized electrons and the polarization of the outgoing nucleons are 
measured (reaction $N(\vec{e},e^\prime \vec{N})$, where $N$ is a nucleon).
The ratio  $G_E/G_M$ is proportional to the 
ratio between the transverse and 
longitudinal nucleon polarizations~\cite{Ndata,Dieterich01}.

In the proton case, the generalization to the nuclear 
medium is based on the reaction  $A(\vec{e},e^\prime \vec{p})B$,
where $B$ is obtained from $A$ by a proton knockout 
in a quasi-elastic reaction~\cite{Dieterich01,JLab-data}.
Although the final proton is
in vacuum, one can still assume that the 
polarization-transfer coefficients carry the information of the bound
proton, since the photon coupling with the proton occurs in
a nuclear environment, and the polarization is conserved.
The medium modification on the electromagnetic form 
factors can then be identified using the double ratio  
$(G_E^\ast/G_M^\ast)/(G_E/G_M)$ between the electric 
and magnetic form factor ratios in medium and in vacuum.
If medium modifications are small, 
the double ratio is close to unity.
Otherwise, one can have an enhancement or a suppression 
of $(G_E/G_M)$ due to medium modifications.

Measurements of the proton electromagnetic form factor
double ratio were performed 
already at MAMI~\cite{Dieterich01} and 
Jefferson Lab~\cite{JLab-data} using 
the $^4$He$(\vec{e},\vec{p})^3$H reaction (proton in medium).
The data show signs of medium modifications,
but the small magnitude of the effect 
may be a consequence of the averaged low density of the $^4$He system 
compared with the 
normal nuclear density 
$\rho_0 = 0.15$ fm$^{-3}$~\cite{MediumDR,Medium1}.
The extension of the method to the neutron 
is planned for the future 
($^4$He$(\vec{e},\vec{n})^3$He reaction)~\cite{PAC35}. 
In progress at MAMI are knockout experiments 
of the proton on $^{14}$C~\cite{MAMI2021,Kolar21}.

The new experiments motivated further theoretical 
studies of the medium modification of the 
electromagnetic form factors,
not only for the proton, but also for the neutron 
based on different frameworks~\cite{Medium1,Calculations1,Calculations2}.
Anticipating future experimental developments, 
we calculate the double ratios for 
all baryon octet members for different values 
of the nuclear density. 
We consider in particular $\rho=0.5 \rho_0$ and $\rho=\rho_0$.

In our calculations of the octet baryon electromagnetic form factors, 
we use the covariant spectator quark model~\cite{CSQM1}.
In the covariant spectator quark model the 
baryons are described as systems of three constituent quarks,
where two of the quarks are on-mass-shell
and can be regarded as an effective diquark, 
and the off-shell quark interacts with the 
electromagnetic fields 
in relativistic impulse approximation~\cite{CSQM1,Octet}.
The electromagnetic structure of the quark is 
parametrized using a vector meson dominance (VMD) form,
which includes contributions of vector meson poles 
depending on the $SU(3)$ channel
(isovector, isoscalar and strange quark)~\cite{Medium1,Octet}.
The radial wave functions are parametrized 
in terms of two momentum range scales according to 
the number of strange quarks 
($SU(3)$ symmetry breaking)~\cite{MediumDR,Octet}.
To determine the parameters of the radial wave functions, 
we extend the model to the lattice QCD regime, in a region 
where meson cloud effects are suppressed (large pion mass region).
The parameters are then adjusted to the lattice QCD data
(details can be found in Refs.~\cite{MediumDR,Medium1,Octet,Lattice}).
In addition to the valence quarks, we consider also 
contributions associated to pion cloud dressing of the baryon cores
determined by the cloudy bag model~\cite{Thomas84}, 
using the $SU(3)$ pion-baryon interaction~\cite{KKCBM}, 
with a $Q^2$-dependence calibrated by the physical data 
for the nucleon and by the hyperon magnetic moments~\cite{Medium1,Octet}.

The generalization to the symmetric nuclear matter
is performed with the assistance of the 
quark-meson coupling (QMC) model~\cite{Saito07},
which has been successfully applied to the study 
of baryons and mesons in nuclei and nuclear medium.
The QMC model is used to estimate the effective masses 
of the baryons and mesons in medium in terms of the nuclear density, 
and furthermore to calculate the effect of those modifications 
on the quark currents and 
radial wave functions in medium~\cite{MediumDR,Medium1}.
The second step is to infer the effect of the medium 
on the pion cloud contributions, which is 
a consequence of the modification of the 
pion-baryon coupling constants in medium~\cite{Medium1}.
The electromagnetic form factors in medium 
are then obtained combining the valence quark 
and pion cloud contributions.

\section{Results and Conclusions}

\begin{figure*}[t]
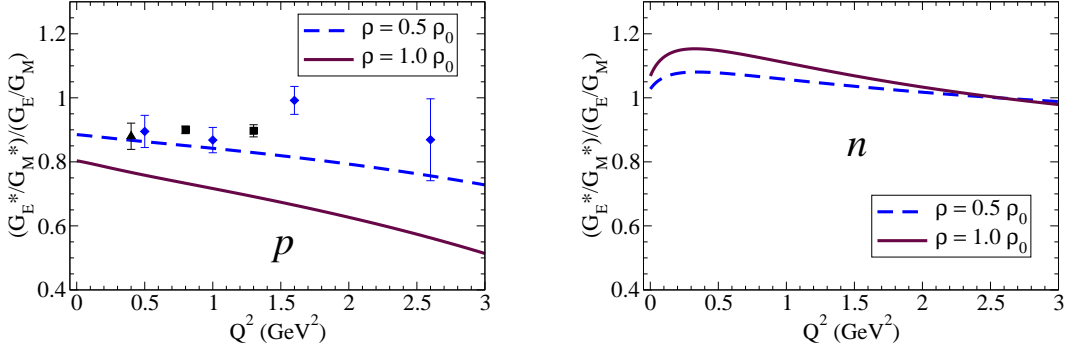

\centerline{
\mbox{
\includegraphics[width=2.5in]{ProtonDRatio.eps} \hspace{1.cm}
\includegraphics[width=2.5in]{NeutronDRatio.eps} }}
\caption{\footnotesize{
Electromagnetic form factor doubles ratios in symmetric nuclear matter 
calculated for proton and neutron.
Experimental data for proton are 
from Refs.~\cite{Dieterich01,JLab-data}.
}}
\label{figNucleon}
\end{figure*}

We calculated the double ratios to all members 
of the baryon octet in symmetric nuclear matter for nuclear densities 
of 0.5$\rho_0$ and $\rho_0$ (normal nuclear density)~\cite{MediumDR}.
The method can be extended in the future to higher densities,
and be applied in studies of heavy-ion collisions,
neutron stars and compact stars.  

\begin{figure*}[t]
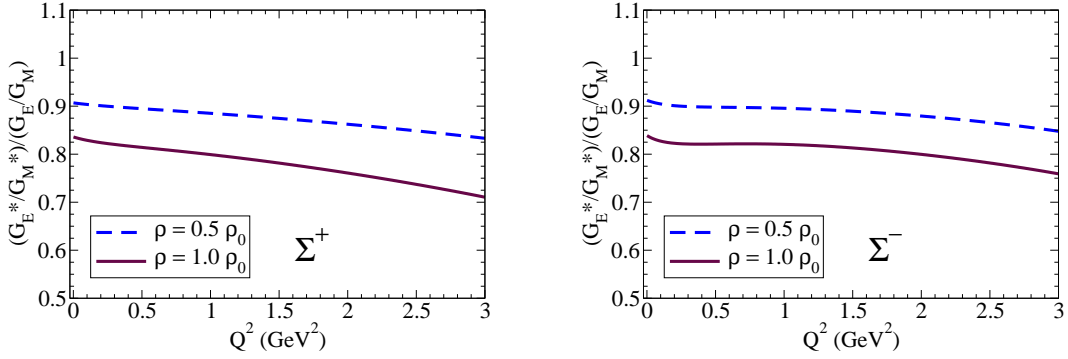

\centerline{
\mbox{
\includegraphics[width=2.5in]{SigmaP-DRatio.eps} \hspace{1.cm}
\includegraphics[width=2.5in]{SigmaM-DRatio.eps} }}
\caption{\footnotesize{
Electromagnetic form factor doubles ratios 
in symmetric nuclear matter calculated for $\Sigma^+$ 
and $\Sigma^-$.
}}
\label{figSigma}
\end{figure*}

In the calculations, we use  covariant constituent quark model
and the quark-meson-coupling model to take 
into account  medium modifications 
on the valence quarks and on the pion cloud contributions.
Our estimates of the proton and neutron, 
and the $\Sigma^+$ and $\Sigma^-$ double ratios, 
are presented in Figs.~\ref{figNucleon}
and \ref{figSigma}, respectively.

In the case of the proton and $\Sigma^\pm$, one 
observes a clear suppression of the ratio $G_E/G_M$ in the nuclear medium. 
The suppression increases with the nuclear density.
In our formalism this suppression is mainly a consequence 
of the dominance of the valence quark effects 
and the properties of the VMD parametrization 
of the quark currents~\cite{MediumDR}.

As for the neutron, we predict an enhancement of the ratio below 2.5 GeV$^2$.
This effect is also predicted by other groups~\cite{Calculations2}.
Our prediction may be tested in a near 
future at Jefferson Lab~\cite{PAC35}.

\begin{figure*}[t]
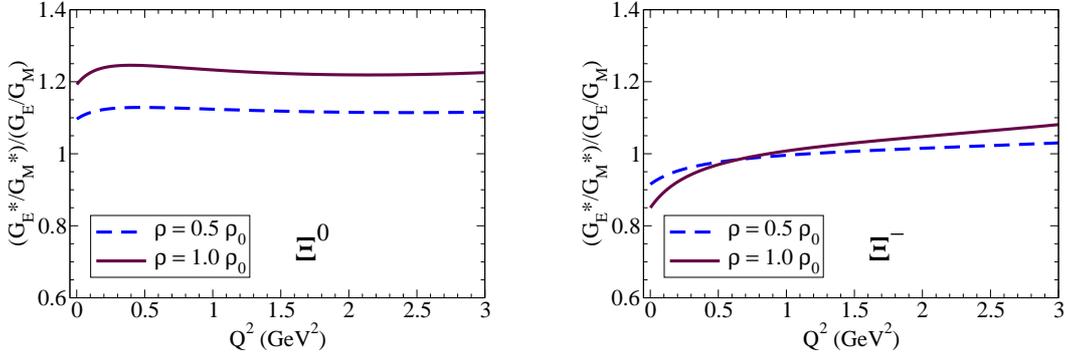

\centerline{\vspace{0.2cm}  }
\centerline{
\mbox{
\includegraphics[width=2.5in]{Xi0-DRatio.eps} \hspace{1.cm}
\includegraphics[width=2.5in]{XiM-DRatio.eps} }}
\caption{\footnotesize{
Electromagnetic form factor doubles ratios 
in symmetric nuclear matter calculated for $\Xi^0$ 
and $\Xi^-$.
}}
\label{figXi}
\end{figure*}

The results for the cascade $\Xi^0$ and $\Xi^-$, 
are presented in Fig.~\ref{figXi}.
From the figure, we can conclude that the $Q^2$-dependence is weak,
as expected, since the systems are dominated by strange quarks.
The results for the $\Lambda$ and $\Sigma^0$ 
and more detailed discussions about the present 
results can be found in Ref.~\cite{MediumDR}.

Within the scope of the PANIC 2021 conference, 
we also emphasize that the present formalism can be used 
to calculate the effective electromagnetic 
form factor $|G(q^2)|$ of baryons at very large $q^2$,
in the timelike region~\cite{HyperonFF}.



\vspace{.3cm}

{\bf Acknowledgments:}  
This work was supported by FAPESP, Brazil,
Process No.~2017/02684-5, Grant No.~2017/17020-BCO-JP (GR),
Process No.~2019/00763-0 (KT) and partially founded by INCT-FNA,
Brazil, Process No.~64898/2014-5, and
CNPq-Brazil, Process No.~401322/2014-9, 
No.~308025/2015-6 (JPBCM), No.~308088/2015-8, No.~313063/2018-4, 
and No.~426150/2018-0 (KT).


\end{document}